\documentclass[pra,twocolumn,superscriptaddress,floatfix]{revtex4-1}
\usepackage[utf8]{inputenc}
\usepackage{graphicx}
\usepackage{graphics,bm}
\usepackage{amssymb}
\usepackage{color}
\usepackage{palatino}
\usepackage{nicefrac}
\usepackage{bm}
\usepackage{mathrsfs}
\usepackage{amsmath}
\usepackage{amsfonts}
\usepackage{times,txfonts}
\usepackage{amsmath}
\usepackage{dsfont}
\usepackage{physics}
\usepackage{bbm}
\usepackage{ulem}

\begin{document}

\title{Precision in estimating Unruh temperature}

\author{D. Borim}
\affiliation{Institute of Physics, Federal University of Goi\'{a}s, POBOX 131, 74001-970, Goi\^{a}nia, Brazil}
\affiliation{Federal Institute of Goi\'{a}s, R. 02 Qd. 10 Lt. 1 to 15, 76600-000, Goi\'{a}s-GO, Brazil}
\author{L. C. C\'{e}leri}
\email{lucas@chibebe.org}
\affiliation{Institute of Physics, Federal University of Goi\'{a}s, POBOX 131, 74001-970, Goi\^{a}nia, Brazil}
\affiliation{Department of Physical Chemistry, University of the Basque Country UPV/EHU, Apartado 644, E-48080 Bilbao, Spain}
\author{V. I. Kiosses}
\affiliation{QSTAR, INO-CNR, Largo Enrico Fermi 2, I-50125 Firenze, Italy}

\begin{abstract}
The goal of quantum metrology is the exploitation of quantum resources, like entanglement or quantum coherence, in the fundamental task of parameter estimation. Here we consider the question of the estimation of the Unruh temperature in the scenario of relativistic quantum metrology. Specifically, we study two distinct cases. First, a single Unruh-DeWitt detector interacting with a scalar quantum field undergoes an uniform acceleration for a finite amount of proper time, and the role of coherence in the estimation process is analyzed. After this, we consider two initially entangled detectors, one of which is inertial while the other one undergoes acceleration. Our results show that the maximum of the Fisher information, thus characterizing the maximum possible precision according to Cramm\'{e}r-Rao bound, occurs only for small accelerations, while it decreases fast when acceleration increases. Moreover, the role of initial coherence ---in the single detector case---, or entanglement ---in the two detectors case---, is to decrease Fisher information. Therefore, under the considered protocol, internal coherence (or entanglement) is not a resource for estimating Unruh temperature. These unexpected results show that a detection of the Unruh effect can be even more challenge than previously thought. Finally, by considering the connection between Unruh effect and Hawking radiation, we discuss how our results can be understood in the context of the estimation of Hawking temperature.
\end{abstract}

\maketitle

\section{Introduction}

Parameter estimation lies in the heart of all sciences and engineering. Therefore, devising schemes for reducing the uncertainty in such a task is a paramount problem, not only for understanding the foundations of science, but also for advancing technology. It is well known that quantum mechanics can be of great help in this endeavor, by providing us quantum resources like coherence and entanglement. Considering these quantum features as a way to improve parameter estimation is the goal of the field known as quantum metrology \cite{Giovannetti2004,Giovannetti2011,Micadei2015}. Several theoretical and experimental developments were achieved so far, among them we can cite quantum illumination \cite{Lloyd2008,Aguilar2019,Barzanjeh2015}, quantum thermometry \cite{Mehboudi2019,Cavina2018} and gravitational effects on quantum matter \cite{Pang2018,Xu2019}, just to mention a few. A general scheme for computing a lower bound on the precision of parameter estimation when noise comes into play was put forward in Ref. \cite{Escher2011}. 

Parallel to these developments, studies of relativistic effects on quantum systems, the carriers of quantum information, had attracted great interest. Such interest is not only due to fundamental reasons, but also due to the possibility of developing quantum technologies working on global scales by using satellites \cite{Rideout2012,Yin2017}. Concerning inertial quantum systems, we mention the discovery that the reduced entropy of a spin system is not observer invariant \cite{Peres2002}, fact that has deep implications for the very definition of spin observable \cite{Celeri2016}. Studies on the Lorentz invariance of entanglement \cite{Gingrich2002,Lamata2006} and on the behaviour of non-locality \cite{Landulfo2009,Landulfo2010} under relativistic motion were also performed.

Concerning non-inertial systems, it was shown that entanglement between two bosonic modes decreases for uniformly accelerated observers \cite{Fuentes2005}. Such decoherence occurs because of the Unruh effect \cite{Unruh1976}, which is a consequence of quantum field theory, pointing out that the very concept of particles is observer dependent.  In short, it states that uniformly accelerated observers in Minkowski spacetime associate a thermal bath of Rindler particles to the Minkowski vacuum (see Ref. \cite{Crispino2008} for comprehensive review on the subject). Interesting, while entanglement is completely destroyed in the case of a scalar field \cite{Adesso2007}, the same does not occur when considering Dirac ones \cite{Alsing2006}. The degradation of more general quantum correlations was also observed in Ref. \cite{Celeri2010}. 

The above cited studies were dedicated to understand how Unruh effect changes the information content of a physical system, or, more specifically, how quantum resources are affected by this effect. Here we are interested in a more practical question regarding the precision that can be achieved in a measurement of the Unruh temperature. Although there are some controversies about the existence of Unruh effect \cite{Narozhny2002,Buchholz2015}, the results of Ref. \cite{Lima2019} seems to theoretically settle the debate by indicating the physical character of the Unruh temperature. However, an experimental verification of such effect is mandatory. An interesting proposal for measuring Unruh temperature using classical electrodynamics was proposed in Ref. \cite{Cozzella2017}, while a simulation of the effect was implemented in Bose-Einstein condensates \cite{Hu2019}. Here we answer the question regarding the precision at which such measurement can be performed, by analyzing the behavior of quantum Fisher information under Unruh effect. We find that small accelerations ---corresponding to small temperatures--- render the estimation of Unruh temperature more precise. Recalling how difficult is to detect a thermal bath caused solely by acceleration (according to Unruh formula, reaching Unruh temperature of order $1 K$, accelerations of order $10^{20}\, m/s^2$ are necessary), our result argues that experimental evidences of existence of Unruh effect are practically even more difficult to be found.

An experimental confirmation of Unruh effect should include, not only a thermal bath detection, but also a verification that the value of measured temperature corresponds to the theoretical prediction. Therefore, precision of temperature measurements should be a concern of any attempt of experimental verification of Unruh effect. The problem of estimating the Unruh temperature was addressed in Refs. \cite{Tian2015,Tian_et_al_2017} from the perspective of a master equation approach. In Ref. \cite{Tian2015}, they concluded that the population measurements are optimal and that the precision of the estimation of the temperature depends on the time evolution and not on the initial state preparation (for long times). In Ref. \cite{Tian_et_al_2017} they have compared the evolution of a static detector and the evolution of an accelerated one and concluded that when the probe atom is initially entangled with a static detector an enhancement in the distinguishability between a static evolution and the accelerated evolution of the probe is observed. By considering two detectors, the authors of Ref. \cite{Wang2014} was able to consider the influence of the detector energy gap and the strength of the interaction between the detector and the scalar field for the precision in the estimation of the Unruh temperature. They also studied the role of entanglement, concluding that it helps improving the precision by increasing Fisher information. 

Here we address the problem of estimation of the Unruh temperature considering two distinct cases, one and two accelerated detectors. In this way we can study the role played by quantum coherences (one detector) and by quantum entanglement (two detectors) in the estimation protocol. We find that these quantum resources actually do not help in the estimation of the Unruh temperature due to the particular nature of the problem. This is a surprising result because it is generally known that entanglement can enhance measurement precision in several situations, including systems affected by thermal noise \cite{Lloyd2008,Aguilar2019,Barzanjeh2015}. These conclusions are reached by analyzing the behavior of quantum Fisher information for both cases. Moreover, we discuss the extension of our results to the estimation of the Hawking temperature near a black hole.

The paper is organized as follows. In the next section we review the Unruh-DeWitt detector and the Unruh effect, thus setting up the system and the notation that will be used in the paper, while Sec. \ref{sec:metrology} is devoted to quantum metrology. Our main results are presented in Secs. \ref{sec:oneD} (one detector) and \ref{sec:twoD} (two detectors). In Sec. \ref{sec:conclusions} we present our final comments, including a discussion of how our results can be extended to the context of Hawking radiation.

\section{Accelerated Detector and the Unruh effect} 
\label{sec:unruh}

In this section we present a very brief description of the Unruh effect, referring the reader to Ref. \cite{Crispino2008} for a complete treatment. Let us start by precisely defining our system. We consider a semiclassical detector interacting with a massless scalar field. Such detectors have a well defined world line in Minkowski spacetime, but their internal degrees of freedom are treated as two level quantum systems \cite{Unruh_Wald84}. The Hamiltonian associated with the internal degrees of freedom is defined as 
\begin{equation}
H_{d}=\omega d^{\dag}d,
\label{HD}
\end{equation}
with $\omega$ representing the detector energy gap. By defining $\left\vert 0\right\rangle$ and $\left\vert 1\right\rangle$ as the ground and excited states of the detector, respectively, the ladder operators $d^{\dag}$ and $d$ can be defined by the relations $d^{\dag}\left\vert 1\right\rangle  = d\left\vert 0\right\rangle = 0$, $d^{\dag}\left\vert 0\right\rangle =\left\vert 1\right\rangle $ and $d\left\vert 1\right\rangle =\left\vert 0\right\rangle $. 

The qubit is coupled to a massless Klein-Gordon field $\phi(x)$ by the interaction Hamiltonian
\begin{equation}
H_{\rm int}(t)=\epsilon(t){\displaystyle\int\limits_{\Sigma_{t}}}\dd^{3} \mathbf{x}
\sqrt{-g} \phi(x)\left[  \varphi(\mathbf{x})d+\varphi^{\ast}(\mathbf{x})d^{\dag
}\right]  ,\label{HI}%
\end{equation}
where $x=(t,\mathbf{x})$ and $\epsilon$, which is a smooth compact support real-valued function, ensures that the detector operates only for a finite amount of proper time $\delta$. $\Sigma_{t=const}$ is a Cauchy surface associated with a timelike isometry and $\mathbf{x}$ are coordinates defined on $\Sigma_{t}$. $g = \det(g_{ab})$ with $g_{ab}$ being the Minkowski (or Rindler) metric. The (smooth compact support complex-valued) function $\varphi(\bf x)$ defines the neighborhood (around the world line) in which the detector interacts with the external field. The total Hamiltonian can then be expressed as $H = H_{0} + H_{\rm int}(t)$, with $H_0 = H_{d} + H_{KG}$, and $H_{KG}$ being the free Klein-Gordon field Hamiltonian (see Ref. \cite{Crispino2008} for more details). 

Working in the interaction picture, we denote by $\ket{\mathbf{\Psi}_{-\infty} }$ the initial state of the system, at the past infinity, and the evolved state will then be given by
\begin{equation}
\ket{\mathbf{\Psi}_t} = T\exp[-i\int_{-\infty}^t \dd t' H_{\rm int} ^I(t')] \ket{\mathbf{\Psi}_{-\infty} },
\end{equation}
where $T$ is the time-ordering operator, while $H_{\rm int}^I (t) = U^{\dagger}_0(t) H_{\rm int} (t) U_0 (t)$, with $U_0 (t) = \exp \qty(-iH_{0}t)$. 

Since the detector operates only for a finite proper time, we must have $\ket{\mathbf{\Psi}_{\infty}} = \ket{\mathbf{\Psi}_{t > \delta}}$ and
\begin{equation}
\ket{\mathbf{\Psi}_{t>\delta}} = T \exp [-i\int \dd^4x\sqrt{-g}\phi(x) (fd + f^*d^{\dagger})] \ket{\mathbf{\Psi}_{-\infty}},
\label{EvoPsi}
\end{equation}
where
$f = \epsilon(t) e^{-i\omega t}\varphi ({\bf x})$. Up to first order, the solution of Eq.~(\ref{EvoPsi}) can be written as
\begin{equation}
\ket{\mathbf{\Psi}_{t>\delta}} = [\mathds{1} - i(\phi(f)\,d + \phi^{\dagger}(f) \,d^{\dagger}) ] \ket{\mathbf{\Psi}_{-\infty}},
\label{Dyson1}
\end{equation}
where $\phi(f)$ is the spacetime average, weighted by the test function $f$, of the field operator \cite{wald94}. It is defined by
\begin{eqnarray}
\phi(f) &\equiv& \int \dd^4 x \sqrt{-g } \phi(x) f \nonumber \\
 & = & i [a((KEf^*)^*)-a^{\dagger}(KEf)]
\label{phi(f)}
\end{eqnarray}
where $a(u^*)$ ($a^{\dagger} (u)$) is the annihilation (creation) operator related to $u$ mode, $K$ is the operator that takes the positive-frequency part of the solutions of Klein-Gordon equation with respect to the timelike isometry, and
\begin{equation}
Ef = \int d^4x'\sqrt{-g(x')} [G_{A}(x, x') - G_{R}(x, x')] f(x'),
\label{Ef}
\end{equation}
with $G_{A}$ and $G_{R}$ being the advanced and retarded Green functions, respectively. Now, if we impose that $\epsilon(t)$ is a very slow-varying function of time compared to the frequency $\omega$ and that $\delta \gg \omega^{-1}$, then $f$ is an approximately positive-frequency function, i.e., $KEf^*\approx 0$, such that $ \phi(f)\approx -i a^{\dagger}(KEf)$ ~\cite{wald94}. With the notation $\lambda \equiv -KEf$, Eq. (\ref{Dyson1}) takes the form
\begin{equation}
\ket{\mathbf{\Psi}_{t>\delta}} = (\mathds{1}+ a^{\dagger}(\lambda)d -  a(\lambda^*)d^{\dagger} ) \ket{\mathbf{\Psi}_{-\infty}}.
\label{first_order_f}
\end{equation}
It is important to notice here that, under the first order approximation, we only consider two distinct processes, in which nothing happens or we observe a transition in the detector (by absorbing/emitting a particle). Although this approximation has physical consequences for quantum correlations between distinct detectors, as discussed in Ref. \cite{Celeri2010}, it will not be crucial in the present article.

\section{Quantum parameter estimation}
\label{sec:metrology}

The paradigmatic problem of metrology is the estimation of a parameter $\xi$ (or set of parameters). This is done in three steps: ($i$) Preparation of the probe state; ($ii$) Interaction of the probe with the system of interest; ($iii$) Measurement of the probe. In the first step we have to prepare a controllable system in a definite, blank state. During step ($ii$) the information about the desired parameter is codified into the state of the probe. This is the codification step. Finally, in the last step, the decoding, we read this information by measuring the probe. By repeating this process several times, we can estimate the value of $\xi$ up to a certain precision. A key quantity in this process is Fisher information, since it bounds the precision (variance) at which $\xi$ can be unbiased estimated by means of the Cramm\'{e}r-Rao inequality \cite{Crammer1945,Rao1945}. The classical Fisher information is defined as
\begin{equation}
 J_\xi=\int_{\mathcal{X}} dx \ \frac{\left[\partial_\xi p(x|\xi)\right]^2}{p(x|\xi)}
 \label{FisClas}
\end{equation}
where $p(x|\xi)$ is the conditional probability of the measurement outcome $x$, given that the desired parameter is $\xi$. $\mathcal{X}$ is the space of all possible outcomes and we used the notation $\partial_\xi \equiv \partial/\partial\xi$. Cramm\'{e}r-Rao inequality reads $\mbox{var}(\xi)\le 1/\sqrt{J_\xi}$, with $\mbox{var}(\xi)$ being the variance associated with $\xi$ \cite{Crammer1945,Rao1945}.

For quantum systems, measurements are described by a set of POVM's (positive-operator valued measure) $\{\Pi_x\}$ and the probabilities are computed applying Born's rule $p(x|\xi)=\mbox{Tr}\qty[\Pi_x \rho_\xi]$, where $\rho_\xi$ is the state of the probe after the codification process. In this scenario, there are several possible sets of POVM's that we can choose and different sets will provide a distinct amount of information about the probe, thus resulting in a different value for the Fisher information. Therefore, the quantum Fisher information is defined as the maximum of $J_{\xi}$ over all possible POVM's. The one that maximizes $J_\xi$ consists of projectors over the eigenstates of the symmetric logarithm derivative (SLD) \cite{ParisFisher}, $L_\xi$, defined by the Lyapunov equation
\begin{equation}
 \partial_\xi \rho_\xi=\frac{L_\xi \rho_\xi+\rho_\xi L_\xi}{2},
 \label{LyapEq}
\end{equation}
whose general solution is given by
\begin{equation}
 L_\xi=2\int_0^\infty dt \ e^{-\rho_\xi t}\partial_\xi\rho_\xi \ e^{-\rho_\xi t}.
\end{equation}

Such maximization procedure results in the quantum Fisher information, $J_\xi^Q$ given by \cite{ParisFisher}
\begin{equation}
J_\xi^{Q} = \mbox{Tr}\qty[\rho_\xi L_\xi^2],
\label{QFisInf}
\end{equation}

Fisher information can be written in a more convenient form if we consider the spectral decomposition of the probe system $\rho_\xi=\sum_jp_j|j\rangle\langle j|$, $0<p_j\leq 1$, $\sum_jp_j=1$. We can then compute the infinitesimal change in the probe state as $\partial_\xi\rho_\xi=\sum_j\partial_\xi p_j|j\rangle\langle j|+p_j|\partial_\xi j\rangle\langle j|+ p_j|j\rangle\langle \partial_\xi j|$. Now, remembering that $\partial_\xi \langle i| j\rangle= \langle \partial_\xi i| j\rangle+ \langle i| \partial_\xi j\rangle=0$, it is possible to show that the quantum Fisher information can be written in the following form \cite{Petz1996,Pires2016}
\begin{equation}
J_{\xi}^{Q} = \sum_{i}\frac{(\partial_\xi p_i)^2}{p_i}+2\sum_{i<j}\frac{2(p_i-p_j)^2}{p_i+p_j}|\langle i|\partial_\xi j\rangle|^2. 
\label{QFIexpr}
\end{equation}
We will refer to the first and second terms as the classical and quantum parts of the Fisher information, respectively. We note that the Cramm\'{e}r-Rao inequality is also valid for $J_{\xi}^{Q}$ \cite{PetzQIT}. This fact allow us to interpret Fisher information as a measure of the amount of information about $\xi$ that was codified in the state $\rho_\xi$.

In the same way, the symmetric logarithm derivative can be written in the form \cite{ParisFisher}
\begin{equation}
 L_\xi=\sum_i \frac{\partial_\xi p_i}{p_i}|i\rangle\langle i| +2\sum_{i\neq j}\frac{p_i-p_j}{p_i+p_j}\langle j|\partial_\xi i\rangle |j\rangle\langle i|.
\end{equation}

After applying a state estimation scheme \cite{PetzQIT} with the optimal POVM to reconstruct $\rho_\xi$, we consider the estimator \cite{ParisFisher}
\begin{equation}
 \mathcal{E}_\xi=\xi\mathds{1}+\frac{1}{J_{\xi}^{Q}}L_\xi.
 \label{OptEst}
\end{equation}
Since $\langle\mathcal{E}_\xi\rangle=\xi$ and $\langle\mathcal{E}_\xi^2\rangle=\xi^2+1/J_{\xi}^{Q}$, $\mathcal{E}_\xi$ saturates  Cramm\'{e}r-Rao inequality.

This characterizes the estimation protocol. We run steps $(i)-(iii)$ several times employing projectors over $L_\xi$ eigenstates in order to obtain data in which the information about the parameter $\xi$ is encoded. The estimator defined in Eq. (\ref{OptEst}) is then used and the value or $\xi$ is computed. Following this protocol we can obtain the optimal estimation of any parameter, including Unruh temperature, which we are interested here.

\section{Quantum Fisher information for accelerated detectors}
\label{sec:oneD}

Let us consider the initial state of the total system (detector and field) as
\begin{equation}
\ket{\mathbf{\Psi}_{-\infty}} = \ket{\psi^d_i} \otimes \ket{0_M},
\label{in_total_state}
\end{equation}
where $\ket{0_M}$ is the field vacuum state in Minkowski spacetime and 
\begin{equation}
\ket{\psi_i^d} = c_0 \ket{0} + c_1\ket{1},
\end{equation}
with $|c_0|^2+|c_1|^2=1$, is the detector initial state. The detector is then accelerated with constant proper acceleration $a$ for the finite amount of proper time $\delta$. Since the detector has a well defined worldline, we choose $\varphi$ as a Gaussian function centered in the classical trajectory \cite{LM2009}
\begin{equation}
\varphi(\mathbf{x})=(\kappa\sqrt{2\pi})^{-3}\exp(-\mathbf{x}^{2}/2\kappa^{2}),
\end{equation}
where $\mathbf{x}$ are the coordinates on the constant time hypersurface and the variance fulfils the condition $\kappa=\mathrm{const}\ll \omega^{-1}$. The detector's worldline is defined by the following equations
\begin{eqnarray}
t(\tau)=a^{-1}\sinh a\tau,\; x(\tau)=a^{-1}\cosh a\tau, \nonumber \\
y(\tau)=z(\tau)=0.
\label{worldline}
\end{eqnarray}
Here, $(t,x,y,z)$ are the Cartesian coordinates in Minkowski spacetime and $\tau$ denotes the detector proper time. Transformations (\ref{worldline}) localize the detector worldline in right Rindler wedge.

By plugging the above expression into Eq. (\ref{first_order_f}) we obtain
\begin{eqnarray}
\ket{\mathbf{\Psi}_{t>\delta}} &=& (c_0\ket{0} + c_1\ket{1})\otimes\ket{0_M} + c_1\ket{0}\otimes a_{W_r}^{\dagger}(\lambda)\ket{0_M} \nonumber \\
&-& c_0\ket{1}\otimes a_{W_r}(\lambda^*)\ket{0_M},
\label{PsiTSol1}
\end{eqnarray}
where the subscript $W_r$ indicates that the creation and annihilation operators act on the (right) Rindler modes. They are connected to the ones acting on Minkoski spacetime by the relations \cite{Unruh_Wald84}
\begin{eqnarray}
a_{W_r}(\lambda^*) &=& \frac{a_M(F_{1\omega}^*) + e^{-\pi \omega/a}a_M^\dag(F_{2\omega})}{(1-e^{-2\pi\omega/a})^{1/2}} \nonumber \\
a_{W_r}^\dag(\lambda)&=&\frac{a_M^\dag(F_{1\omega})+e^{-\pi \omega/a}a_M(F_{2\omega}^*)}{(1-e^{-2\pi\omega/a})^{1/2}}.
\label{creatRI}
\end{eqnarray}
The positive frequency solutions are given by
\begin{eqnarray}
F_{1\omega} &=& \frac{\lambda+e^{-\pi \omega/a}\lambda\circ \zeta}{(1-e^{-2\pi\omega/a})^{1/2}} \nonumber \\
F_{2\omega} &=& \frac{(\lambda\circ \zeta)^* + e^{-\pi \omega/a}\lambda^*}{(1-e^{-2\pi\omega/a})^{1/2}}.
 \label{ModeF}
\end{eqnarray}
Here, $\zeta$ denotes the Rindler wedge reflection isometry $\zeta(t,x,y,z) = (-t,-x,y,z)$. In order to further develop Eq. ~(\ref{PsiTSol1}), we introduce the Klein Gordon inner product $(,)_{KG}$ of the modes (\ref{ModeF})
\begin{equation}
(F_{i\omega},F_{j\omega})_{KG} = \parallel \lambda\parallel^2\delta_{ij},
\end{equation}
with \cite{LM2009,Celeri2010}
\begin{equation}
\parallel \lambda\parallel^2 = \frac{\epsilon^2\omega\delta}{2\pi}e^{-\omega^2\kappa^2}\equiv \mu.
\label{mudef}
\end{equation}
By introducing the Unruh temperature $T=a/2\pi$, Eq. (\ref{PsiTSol1}) becomes
\begin{eqnarray}
\ket{\mathbf{\Psi}_{t>\delta}} &=& (c_0\ket{0} + c_1\ket{1})\otimes\ket{0_M} \nonumber \\ 
&+& c_1\ket{0} \otimes \frac{\mu^{1/2}}{(1- e^{-\omega/T})^{1/2}}\ket{1_{F_{1\omega}}} \nonumber \\
&-& c_0\ket{1}\otimes \frac{\mu^{1/2}e^{-\omega/2T}}{(1-e^{-\omega/T})^{1/2}}\ket{1_{F_{2\omega}}},
\label{PsiTSol2}
\end{eqnarray}
where $\ket{1_{F_{i\omega}}}$ ($i=1,2$) denotes the normalized field state corresponding to the modes (\ref{ModeF}).

We are interested in the Fisher information associated with the detector. Therefore, we define $\rho_{d} = \mbox{Tr}_{\phi}\ket{\mathbf{\Psi}_{t>\delta}}\bra{\mathbf{\Psi}_{t>\delta}}$, with $\mbox{Tr}_{\phi}$ denoting the partial trace over the field degrees of freedom, as the reduced density matrix of the detector. It is given by 
\begin{equation}
\rho_d = \mathcal{N}\left(
 \begin{array}{cc}
 f_\omega|c_0|^2+\mu|c_1|^2 & f_\omega c_0 c_1^* \\
 f_\omega c_0^* c_1 & \mu e^{-\omega/T}|c_0|^2+f_\omega|c_1|^2
 \end{array}
 \right),
\label{FinalStateDetector}
\end{equation}
where $f_{\omega} = 1-e^{-\omega/T}$ and $\mathcal{N} = (f_{\omega} + \mu e^{-\omega/T}|c_0|^2+\mu |c_1|^2)^{-1}$ is the normalization constant. 

Now we ask about how much information the state $\rho_d$ contains about the Unruh temperature, the parameter of our interest here. This is obtained by computing the Fisher information associated with this state. From here on we set $\omega=1$, implying that the Unruh temperature and the acceleration will be measured in units of $\omega$ while the quantum Fisher information is measured in units of $\omega^{-2}$. We also consider $c_0,c_1\in\mathbb{R}$, so we can write $c_0 = \sin\eta$ and $c_1 = \cos\eta$. After diagonalizing state $\rho_d$ it is straightforward to obtain $J_{T}^{Q}$ from Eq. (\ref{QFIexpr}). The final expression is too cumbersome to be shown here and is given in Appendix \ref{app:fisherSD}. Figure \ref{Fig1} shows the behaviour of $J_{T}^{Q}$ as function of acceleration for distinct initial states.

\begin{figure}[h]
\begin{center}
\includegraphics[width=\linewidth]{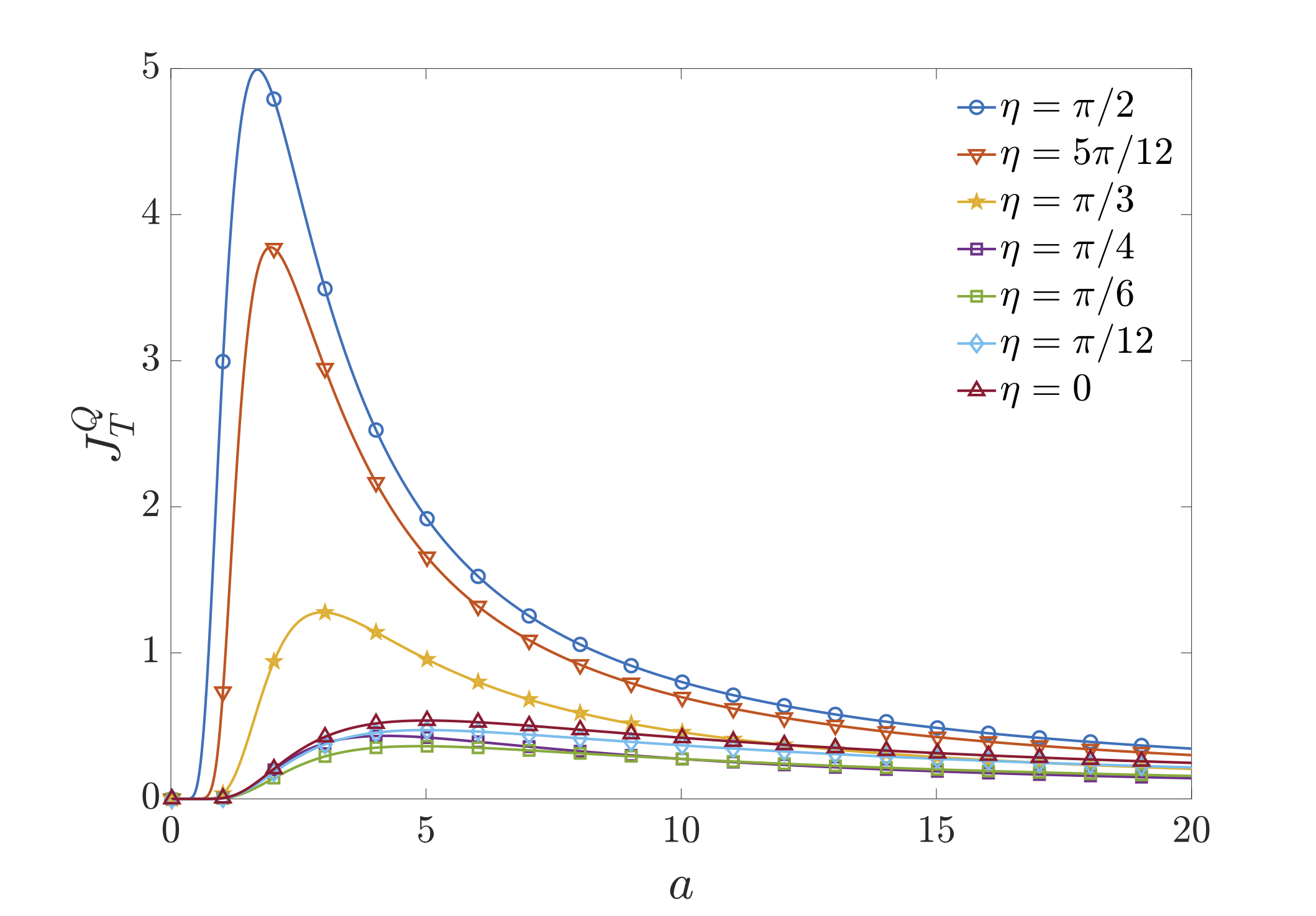} 
\caption{Quantum Fisher information $J_{T}^{Q}$ as function of acceleration $a$ for the single detector case. Distinct lines represent different initial states, as labelled by the values of $\eta$. All values of $J_{T}^{Q}$ are expressed in units of $(10^{-2}\omega^{-2})$.}
\label{Fig1}
\end{center}
\end{figure}

\begin{figure}[h]
\begin{center}
\includegraphics[width=\linewidth]{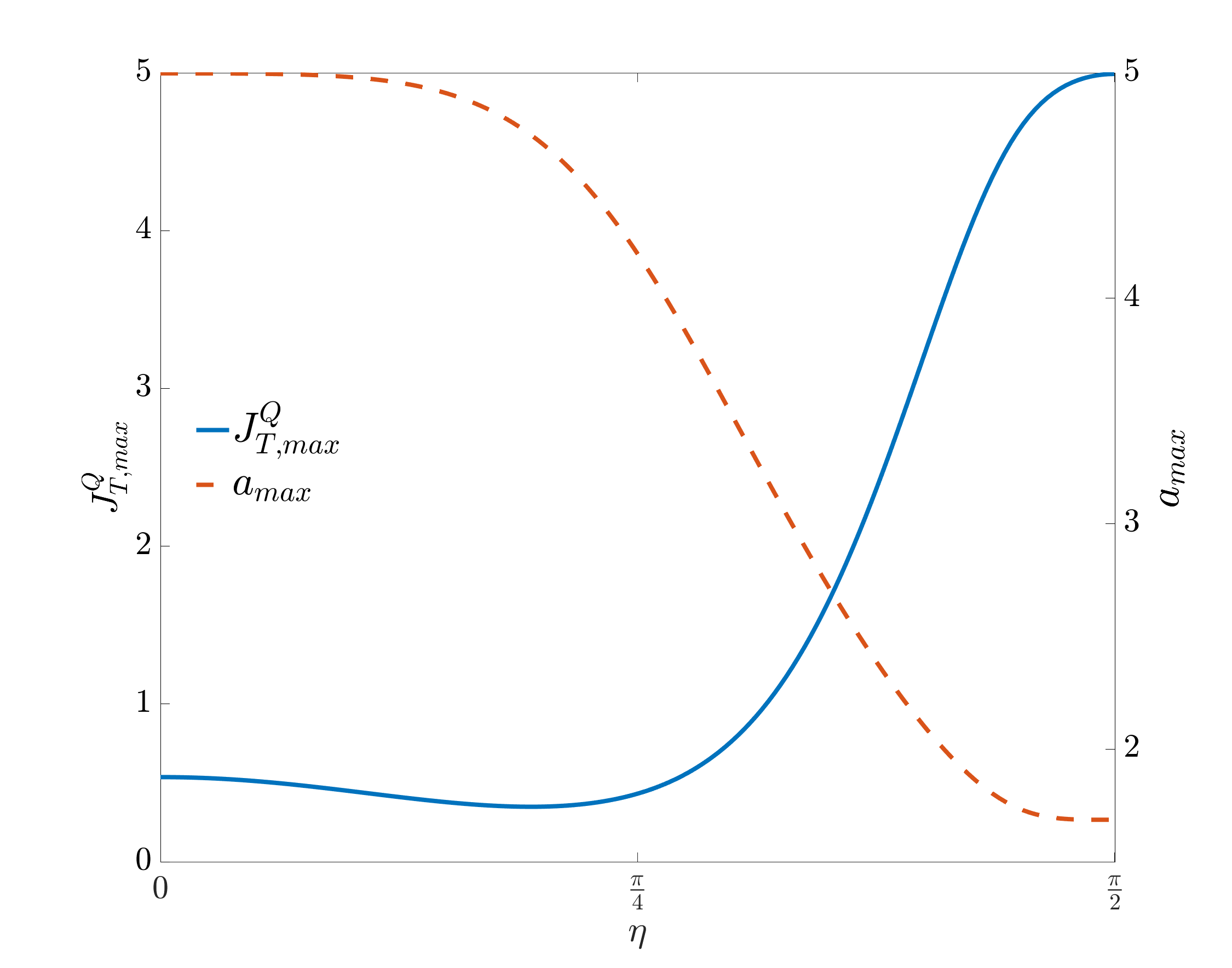} 
\caption{Maximum value of quantum Fisher information $J_{T, max}^{Q}$ and the corresponding acceleration $a_{max}$ as function of $\eta$.}
\label{Figmax1qu}
\end{center}
\end{figure}

We observe a maximum of $J_{T}^{Q}$ for relatively small values of the acceleration, indicating that there is an optimal value of the Unruh temperature that can be measured with high precision. As the acceleration increases (higher temperatures), quantum Fisher information approaches zero asymptotically, thus indicating that the minimum possible variance increases. This behavior, which holds for all values of $\eta$, can be understood as follows. In each run of the experiment, we prepare the initial state of the detector (the probe system). After coupling it to the field, the detector is uniformly accelerated during a finite amount of time. This is the codification process, where the information about Unruh temperature is recorded in the detector's state. The final step, the decoding, is the measurement (after one click or the end of time $\delta$). By repeating this procedures many times we are able to estimate $T$. For large values of the acceleration the thermal fluctuations increase, which makes the detection in changes in the temperature very difficult. As we can see in Appendix \ref{app:fisherSD}, Fisher information goes to zero, $J_{T}^{Q}\rightarrow 0$, as $a\rightarrow \infty$. In order to address this behavior, in Fig. \ref{Figmax1qu} we show the maximum value of Fisher information, $J_{T, max}^Q$, and the associated acceleration, $a_{max}$, as function of the initial state. Note that the optimal initial state for the estimation of $T$ is also the one with smallest $a_{max}$.

We also observe that the optimal state for temperature estimation (the one with the highest values of $J_{T}^{Q}$) is when $\eta=\pi/2$, i.e. the state $\ket{\psi_i^d} = \ket{0}$. This can be explained considering the interaction between the detector and the external field. If the detector is initially in it's excited state, the interaction with the vacuum can induce an spontaneous decay independently of the acceleration. The probability of observing a detector transition in this case, for very small acceleration, increases due to the mixture of both effects thus resulting in a smaller amount of information about the Unruh temperature codified in the state of the detector.

Finally, we note that $J_{T}^{Q}$ increases with $\mu$ for all values of $\eta$. This is expected since the greater the period of acceleration, the larger the probability of observing a detector transition.

In the next section we consider the role of entanglement in the behavior of quantum Fisher information.

\section{Entanglement and Estimation Precision}
\label{sec:twoD}

Let us consider two identical detectors localized in distinct spacetime regions, meaning that their localization function $\varphi(\mathbf{x})$ do not overlap. One of the detectors will be accelerated, with the trajectory determined by Eq. (\ref{worldline}), while the other one will be kept inertial. We consider the following initial state for the entire system
\begin{equation}
\ket{\mathbf{\Psi}_{-\infty}} = \ket{\psi_i} \otimes \ket{0_{M}},
\label{InitState}
\end{equation}
with $\ket{\psi_i} = \sin\theta \ket{01} + \cos\theta \ket{10}$. We use the notation $\ket{xy} = \ket{x}\otimes\ket{y}$ ($x, y \in \{0,1\}$) to represent the tensor product of the state of the two detectors. The first entry refers to the inertial detector and the second to the accelerated one. We should emphasize that the interaction described by Eq. (\ref{HI}) applies only to the accelerated detector, not to the inertial one. Therefore, Eq. (\ref{first_order_f}) holds in this case with operators $d$ and $d^\dagger$ acting only on second entry of $\ket{\psi_i}$. By applying Eq. (\ref{first_order_f}) to $\ket{\mathbf{\Psi}_{-\infty}}$ we obtain
\begin{eqnarray}
\ket{\mathbf{\Psi}_{t>\delta}} &=& \ket{\mathbf{\Psi}_{-\infty}} + \sin\theta \ket{00}\otimes  a^\dagger_{W_r}(\lambda)\ket{0_{M}}\nonumber \\
&-& \cos\theta\ket{11}\otimes a_{W_r}(\lambda^*)\ket{0_{M}}.
\label{psi_f_1}
\end{eqnarray}

Considering Eqs. (\ref{creatRI}) to (\ref{mudef}), we can write
\begin{eqnarray}
\ket{\mathbf{\Psi}_{t>\delta}} &=& \frac{1}{\sqrt{\mathcal{N}}} \left( \frac{\sin\theta\mu^{1/2}}{(1-e^{-\omega/T})^{1/2}}\ket{00}\otimes\ket{1_{F_{1\omega}}}\right. \nonumber \\
&+& \sin\theta\ket{01}\otimes\ket{0_M} + \cos\theta\ket{10}\otimes\ket{0_M} \nonumber \\
&-& \left. \frac{\cos\theta e^{-\frac{1}{2}\frac{\omega}{T}}\mu^{1/2}}{f_{\omega}^{1/2}}\ket{11}\otimes\ket{1_{F_{2\omega}}} \right).
\end{eqnarray}
where $\mathcal{N} = 1+ f_{\omega}^{-1}\left(\sin^2\theta \mu + \cos^2\theta e^{-\omega/T}\mu\right)$.

By tracing out the field degrees of freedom, we get the following reduced density matrix from the evolved detectors (in the basis $\{|00\rangle,|01\rangle,|10\rangle,|11\rangle\}$)
\begin{equation}
 \rho_{dd} = \frac{1}{\mathcal{N}}
 \left(
 \begin{array}{cccc}
 \frac{\sin^2\theta \mu}{f_{\omega}} & 0 & 0 & 0\\
 0 & \sin^2\theta & \cos\theta \sin\theta & 0\\
 0 & \sin\theta \cos\theta & \cos^2\theta & 0\\
 0 & 0 & 0 & \frac{\cos^2\theta e^{-\omega/T} \mu}{f_{\omega}}
 \end{array}
 \right) 
\label{GenFinState}
\end{equation}

We want to look for a relation between optimal conditions to estimate Unruh temperature and the entanglement present between our detectors. The density operator shown in Eq. (\ref{GenFinState}) can be written in a diagonal form $\rho_{dd} = \mbox{diag}(\alpha,\beta,\gamma,0)$ in the basis $\{\cos\theta \ket{10} + \sin\theta\ket{01},\ket{00},\ket{11},-\sin\theta\ket{10} + \cos\theta\ket{01}\}$, where
\begin{align*}
\alpha  &  =\frac{f_{\omega}}{f_{\omega}+\mu(\sin^2\theta + e^{-\omega/T}\cos^2\theta)},\\
\beta  &  =\frac{\mu\sin^2\theta}{f_{\omega}+\mu(\sin^2\theta+e^{-\omega/T}\cos^2\theta)},\\
\gamma  &  =\frac{\mu e^{-\omega/T}\cos^2\theta }{f_{\omega}+\mu(\sin^2\theta+e^{-\omega/T}\cos^2\theta)}.
\end{align*}

It is interesting to note here that the eigenbasis of $\rho_{dd}$ does not depend on the temperature $T$, which is the parameter we want to estimate. Therefore, the quantum contribution to the Fisher information, given by the second term in Eq. (\ref{QFIexpr}), vanishes and we got just the classical term of $J_{T}^{Q}$. The complete expression for the Fisher information for this case is shown in Appendix \ref{app:fisherDD}. Figure \ref{Fig2} shows $J_{T}^{Q}$ as a function of the acceleration $a$ for several values of $\theta$, which quantifies the initial entanglement. The same pattern obtained for the single detector case is observed here, with the difference that now distinct curves represents different amounts of entanglement, while in the single detector case, they describe the amount of quantum coherence in the initial eigenbasis of the detector. Although we can have relatively high values of the Fisher information for all values $\theta$, the optimal state for temperature estimation is the separable one $\ket{\psi_i} = \ket{10}$. Also, for high temperatures, estimation becomes harder. In fact, we have $J^Q_{T}\rightarrow 0$ as $a\rightarrow 0$ (see Appendix \ref{app:fisherDD}).The reason for this behavior is the same presented in single detection case. The maximium value of Fisher information and the associated acceleration as function of the initial state is shown in Fig. \ref{Figmax2qu}.

\begin{figure}[h]
\begin{center}
\includegraphics[width=\linewidth]{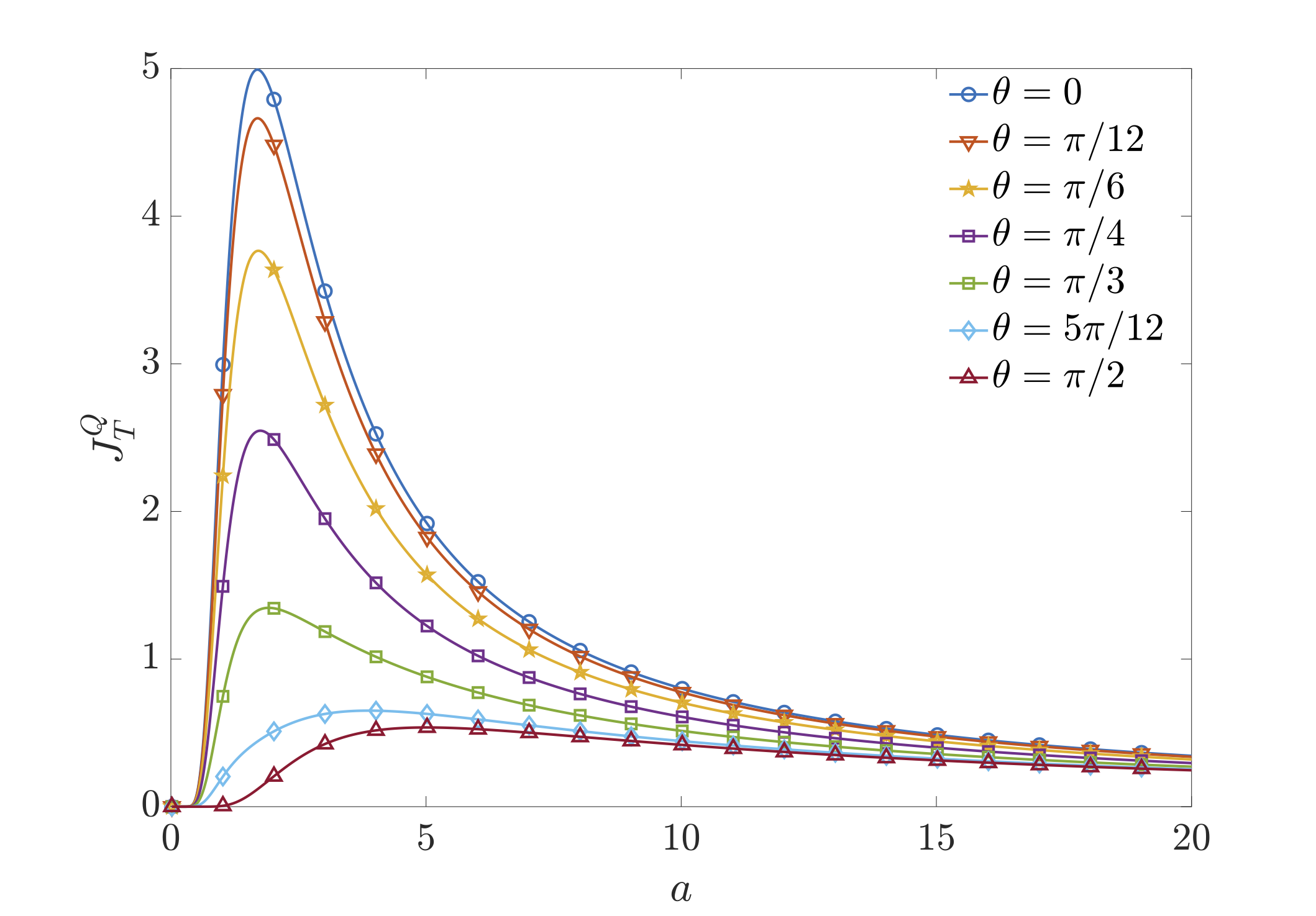}
\caption{Quantum Fisher information $J_{T}^{Q}$ as function of the acceleration $a$ for distinct initial states. Again, we set $\omega = 1$.}
\label{Fig2}
\end{center}
\end{figure}

\begin{figure}[h]
\begin{center}
\includegraphics[width=\linewidth]{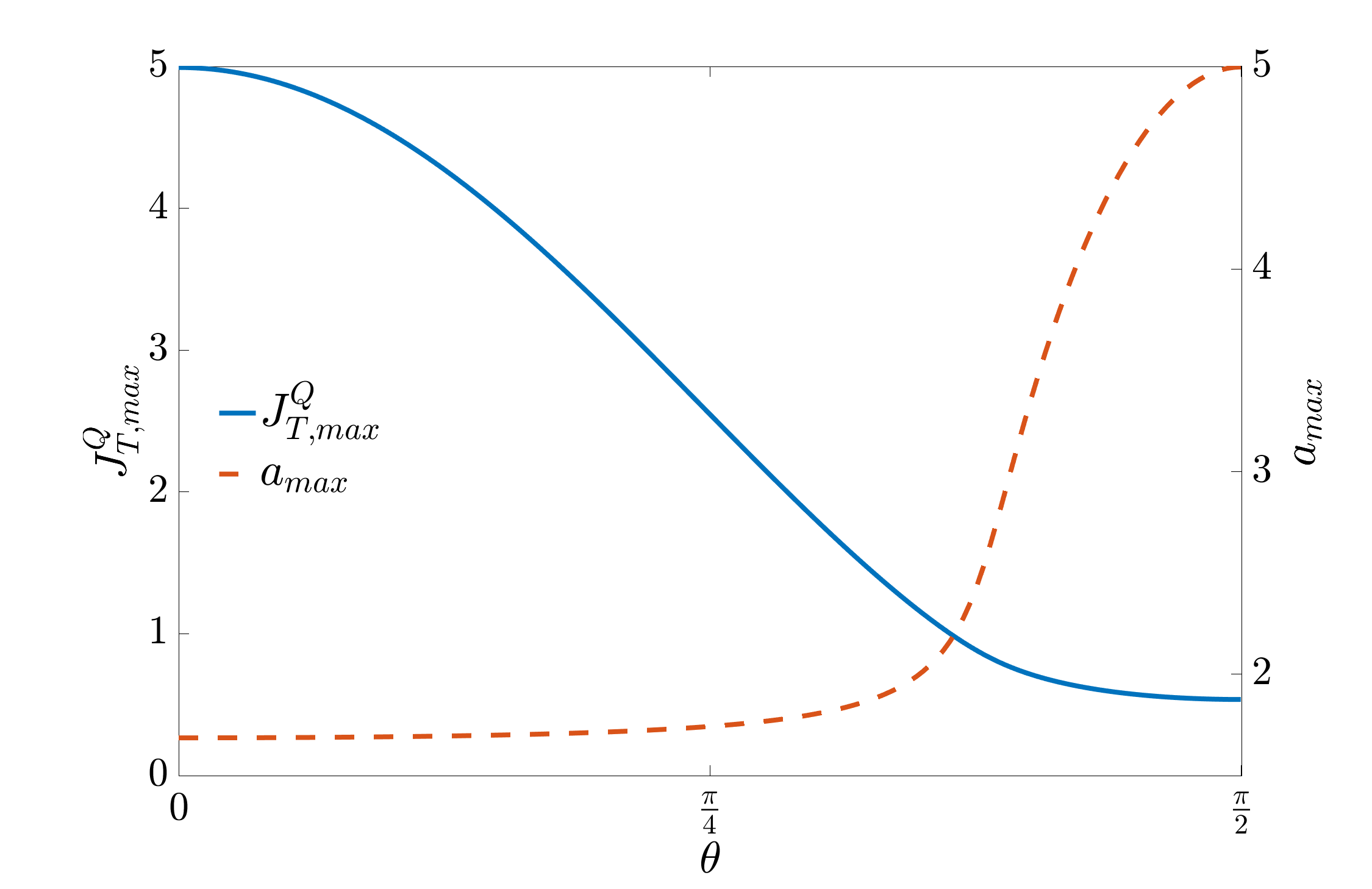}
\caption{Maximum value of quantum Fisher information $J_{T, max}^{Q}$  and the associated acceleration $a_{max}$ as function of $\theta$.}
\label{Figmax2qu}
\end{center}
\end{figure}

\begin{figure}[h]
\begin{center}
\includegraphics[width=\linewidth]{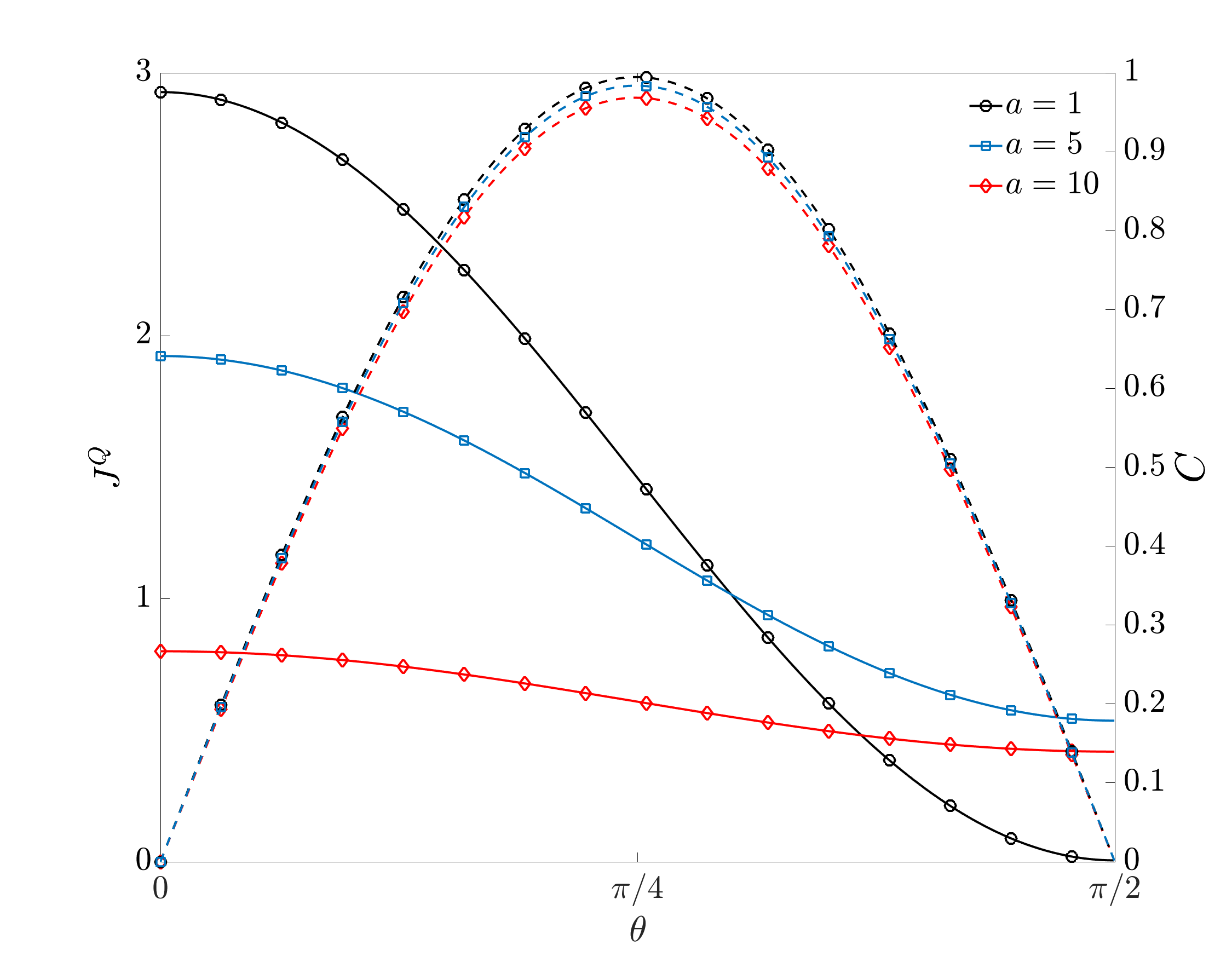} 
\caption{Quantum Fisher information $J_{T}^{Q}$ as function of $\theta$ for different values of the acceleration $a$. Dashed lines represent the concurrence and solid lines $J_{T}^{Q}$. Black line holds for $a=1$, green for $a=5$, blue for $a=10$. For all curves we set $\mu=0.01$.}
\label{Fig3}%
\end{center}
\end{figure}

In order to better understanding the role played by the initial entanglement in the estimation of Unruh temperature, we investigate the relation between Fisher information and the concurrence of $\rho_{dd}$, which is given by
\begin{equation}
 C(\rho_{dd})=\max\{0,\sqrt{\lambda_1}-\sqrt{\lambda_2}-\sqrt{\lambda_3}-\sqrt{\lambda_4}\},
\end{equation}
where $\lambda_i$, $i\in\{1,2,3,4\}$, are eigenvalues of the operator $\rho_{dd}(\sigma_y \otimes \sigma_y)\rho_{dd}^*(\sigma_y \otimes \sigma_y)$, $\sigma_j$ ($j\in\{x,y,z\}$) are the Pauli matrices. The eigenvalues are ordered such that $\lambda_1\ge\lambda_2\ge\lambda_3\ge\lambda_4$. As we can observe in Fig. \ref{Fig3}, for small values of acceleration, $J_{T}^{Q}$ increases as $\theta$ decreases implying that smaller amounts of entanglement is better for the estimation of temperature, contrary to what had been suggested in literature \cite{Wang2014}. This is something we expected since the presence of entanglement tends to decrease the purity of the detector state. This can be clearly seen from the fact that the maximum of $J_{T}^{Q}$ occurs for $\theta = 0$ (which corresponds to initial separable state $\ket{10}$), while the maximum of entanglement occurs for $\theta=\pi/4$. Note that the decoherence induced by the Unruh effect \cite{LM2009,Celeri2010} destroys the entanglement (and all the quantum correlations) and, consequently, the information about $T$ contained in the final state of the detectors decreases. The final state of the detectors contains only classical correlations (as a consequence of the first order approximation) \cite{Celeri2010}, which cannot be used in order to get any quantum advantage for parameter estimation.

Note that there is an asymmetry in our system. While the state $\ket{10}$ works well for the estimation protocol, state $\ket{01}$ does not. This reflects the fact that the inertial detector does not interact with the external field while the accelerated one does, even in the absence of acceleration. Therefore, there is a finite probability that the state $\ket{01}$ decays to $\ket{00}$, while the state $\ket{10}$ can only flip to $\ket{11}$ if there is acceleration. So, starting the experiment with the detector in state $\ket{10}$ ($\theta = 0$) improves the estimation of the Unruh temperature. 

Finally, we note that $J_{T}^{Q}$ increases with $\mu$. This is expected since the greater the period of acceleration, larger the probability of observing a detector click, for small enough acceleration values.

\section{Discussions}
\label{sec:conclusions}

We analyzed the precision of estimation of Unruh temperature using Fisher information for two distinct cases: ($i$) single accelerated detector and ($ii$) one inertial and one accelerated detector. Detectors are modeled as Unruh-DeWitt two-level semiclassical systems coupled to an external field. In the single detector case we observed the maximum the Fisher information associated with Unruh temperature for relatively small values of acceleration ($\le 5 \omega$). When the acceleration increases, Fisher information decreases, asymptotically approaching zero and the precision of estimation gets poorer due to the increase of thermal fluctuations. It gets harder to distinct small differences in the temperature. We conclude that the unexcited state is the optimal initial state for temperature estimation, thus showing that internal coherence is not a resource in this case. This holds because the excited state has a decay probability independent of acceleration and the mixture of Unruh and spontaneous decay effects results in a smaller amount of information about Unruh temperature codified in its final state. We also observed that the maximum of Fisher information increases with increasing the interaction time $\delta$, but the decaying behavior does not change.

The case of one inertial and one accelerated detector was studied in order to analyze the role of the entanglement in the precision in the measurement of Unruh temperature. The maximum of Fisher information is again observed only for small values of acceleration ($\le 5 \omega$ again), fast decreasing for high acceleration values. Our results show that entanglement plays no role in our protocol. The optimal initial state for temperature measurement is the separable, unexcited state $|10\rangle$. This is explained, as in the single detector case, by the fact that excited detector has a decay probability that has no relation with Unruh effect, so a decaying event would be less informative about Unruh temperature then a click of the unexcited state. The same behavior with respect to $\delta$ is also observed here.

One way to increase the precision in the estimation of a parameter is by using quantum resources, like coherence and entanglement. It is true, in general, that such resources indeed help and their consumption is translated into higher precision. What we showed here is the opposite, that quantum resources do not help. Our results demonstrate that quantum metrology does not help in this case, and that a reliable measurement of the Unruh effect can be done only for small accelerations, thus increasing the challenge for an experimental implementation.

It is interesting to observe that our results can be translated to the estimation of Hawking radiation for accelerating detectors very close to the event horizon of a Schwarzschild black hole. To understand how this can be done, let us consider that the accelerated detector (and non-rotating) is located at a fixed radius $r$, just outside the event horizon $r_S$ of the black hole.  For $r$ very close to $r_S$ (the Schwarzschild radius), the acceleration $a$ of the detector should be very large, otherwise it cannot stand still. This implies that the timescale $t=1/a$ (remember that $c=1$) will be much smaller than $r_S$. Therefore, at these scales, the spacetime curvature is negligible and we expect that the accelerating detector experiences the Unruh effect, as discussed in the previous sections.

The difference is that now, due to gravity, the outgoing radiation (the thermal bath modes) will be red-shifted as they move away from the black hole. Given that the norm $\chi$ of the time-translation Killing field (defining the positive- and negative-frequency modes) is $1/a$, the ratio between the temperatures as measured by static detectors at two different radii is simply
\begin{equation}
\frac{T_{r_1}}{T_{r_2}} = \frac{\chi_2}{\chi_1}.
\end{equation}
At infinity it holds that $\chi=1$, so if we assume $\chi_1$ at infinity (the inertial detector), the temperature at infinity will be 
\begin{equation}
T_{r_1}=\frac{\chi_2 a}{2\pi}=\frac{\kappa}{2\pi},
\end{equation}
where we have introduced the surface gravity $\kappa$. For the specific case of Schwarzschild black hole it holds that $\kappa=1/2r_S$ or $\kappa=1/4m$, where $m$ is the black hole mass.

%
%
\vspace{0.2cm}
\noindent
{\bf Acknowledgments} --- DB and LCC thanks the Brazilian Agencies CNPq, FAPEG, and the Brazilian National Institute of Science and Technology of Quantum Information (INCT/IQ) for the financial support. This study was financed in part by the Coordena\c{c}\~{a}o de Aperfeiçoamento de Pessoal de N\'{i}vel Superior - Brasil (CAPES) - Finance Code 001. LCC also acknowledges support from Spanish MCIU/AEI/FEDER (PGC2018-095113-B-I00), Basque Government IT986-16, the projects QMiCS (820505) and OpenSuperQ (820363) of the EU Flagship on Quantum Technologies and the EU FET Open Grant Quromorphic and the U.S. Department of Energy, Office of Science, Office of Advanced Scientific Computing Research (ASCR) quantum algorithm teams program, under field work proposal number ERKJ333.

\appendix

\section{Fisher information for a single detector}
\label{app:fisherSD}

The evolved density operator $\rho_d$ for the single detector case is shown in Eq. (\ref{FinalStateDetector}). If $\eta = 0$ or $\eta = \pi/2$, then $\rho_d$ is already diagonal and we can directly apply Eq. (\ref{QFIexpr}) to obtain the following results for the quantum Fisher information
\begin{eqnarray}
 J_T^Q(\eta=\pi/2)=\frac{e^{-\omega/T}\mu\omega^2}{T^4f_\omega(f_\omega + e^{-\omega/T}\mu)^2},\\
 J_T^Q(\eta=0)=\frac{e^{-2\omega/T}\mu\omega^2}{T^4f_\omega(f_\omega + \mu)^2}.
\end{eqnarray}
For all other values, $\eta \in \left(0,\pi/2\right)$, we can write $\rho_{d}$ in the diagonal form $\rho_{d}=\mbox{diag}(1/2-\lambda,1/2+\lambda)$, where $\lambda$ is given by
\begin{equation}
 \lambda=\frac{\sqrt{\lambda_{1}+\lambda_{2}+\lambda_{3}}}{\lambda_{4}},
\end{equation}
where 
\begin{align*}
&\lambda_{1} = e^{-2\omega/T}(8+\mu(4+3\mu))-2 e^{-\omega/T}(8+\mu^2)+(8+\mu(-4+3\mu)),\\
&\lambda_{2}=4\mu f_\omega(-2+\mu+e^{-\omega/T}(2+\mu))\cos(2\eta),\\
&\lambda_{3}=\mu (1+e^{-\omega/T})(-4+\mu+e^{-\omega/T}(4+\mu))\cos(4\eta),\\
&\lambda_{4}=2\sqrt{2}\left[(e^{-\omega/T}+1)\mu + f_{\omega}(2 + \mu\cos(2\eta))\right].
\end{align*}
The eigenbasis of the density operator is $\{V_1,V_2\}$, with
\begin{equation*}
 V_1=1/N_1\left(\frac{1}{4f_\omega\sin(2\eta)}(\lambda_{5}-\sqrt{2}\sqrt{\lambda_{1}+\lambda_{2}+\lambda_{3}}),1\right),
\end{equation*}
and 
\begin{equation*}
 V_2 = \frac{1}{N_2}\left(\frac{1}{4f_\omega\sin(2\eta)}(\lambda_{5}+\sqrt{2}\sqrt{\lambda_{1}+\lambda_{2}+\lambda_{3}}),1\right).
\end{equation*}
Where 
\begin{align*}
\lambda_{5}=-2e^{-\omega/T}(\mu-(2+\mu)\cos(2\eta))+2(\mu+(-2+\mu)\cos(2\eta)),
\end{align*}
and $N_1$ and $N_2$ are the normalization constants
\begin{eqnarray}
 N_1=\sqrt{\left(\frac{(\lambda_{5}-\sqrt{2}\sqrt{\lambda_{1}+\lambda_{2}+\lambda_{3}})}{4f_\omega\sin(2\eta)}\right)^2+1}\nonumber\\
 N_2=\sqrt{\left(\frac{(\lambda_{5}+\sqrt{2}\sqrt{\lambda_{1}+\lambda_{2}+\lambda_{3}})}{4f_\omega\sin(2\eta)}\right)^2+1}
\end{eqnarray}

A direct application of Eq. (\ref{QFIexpr}) lead us to the quantum Fisher information.

\section{Fisher information for two detectors}
\label{app:fisherDD}

The evolved state of both detectors, $\rho_{dd}$, is given by Eq. (\ref{GenFinState}). As in the case of a single detector, the simple cases where $\theta = 0$ and $\theta = \pi/2$, $\rho_{dd}$ is already diagonal and it is straightforward to apply Eq. (\ref{QFIexpr}) to obtain the quantum Fisher information
\begin{eqnarray*}
 J_T^Q(\theta=0) &=& \frac{e^{-\omega/T}\mu\omega^2}{T^4f_\omega(f_\omega + e^{-\omega/T}\mu)^2},\\
 J_T^Q(\theta=\pi/2) &=& \frac{e^{-2\omega/T}\mu\omega^2}{T^4f_\omega(f_\omega+\mu)^2}.
\end{eqnarray*}

In the general case, $\theta \in \left(0,\pi/2\right)$, the density operator of both detectors can be written as $\rho_{dd} = \mbox{diag}(\alpha,\beta,\gamma,0)$ in the basis $\{\cos\theta \ket{10} + \sin\theta\ket{01},\ket{00},\ket{11},-\sin\theta\ket{10} + \cos\theta\ket{01}\}$, where
\begin{eqnarray*}
\alpha &=& \frac{f_{\omega}}{f_{\omega}+\mu(\sin^2\theta + e^{-\omega/T}\cos^2\theta)},\\
\beta &=& \frac{\mu\sin^2\theta}{f_{\omega}+\mu(\sin^2\theta+e^{-\omega/T}\cos^2\theta)},\\
\gamma &=& \frac{\mu e^{-\omega/T}\cos^2\theta }{f_{\omega}+\mu(\sin^2\theta+e^{-\omega/T}\cos^2\theta)}.
\end{eqnarray*}

Now, we note that the eigenbasis of the density operator does not depend on $T$, which makes easier the application of Eq. (\ref{QFIexpr}) for computing the quantum Fisher information, that turns out to be written, in this case, as
\begin{equation*}
 J_T^Q=\frac{(\partial_T \alpha)^2}{\alpha}+\frac{(\partial_T \beta)^2}{\beta}+\frac{(\partial_T \gamma)^2}{\gamma}.
\end{equation*}
By computing the derivatives, the final form for the quantum Fisher information is
\begin{widetext}
\begin{equation*}
 J_T^Q=\frac{e^{-\omega/T}\mu\omega^2(e^{-\omega/T}(4(1-\cos(2\theta))+\mu(-1+\cos(4\theta)))+4(1+\cos(2\theta))+\mu(1-\cos(4\theta)))}{2T^4f_\omega(e^{-\omega/T}(-2+\mu(1+\cos(2\theta)))+2+\mu(1-\cos(2\theta)))^2}.
\end{equation*}
\end{widetext}


\end{document}